\documentclass[twocolumn,showpacs,aps]{revtex4}
\usepackage[dvips]{graphicx}
\usepackage{bm}
\usepackage{mathrsfs}
\usepackage{lipsum}
\usepackage{graphicx,epsfig}
\usepackage{psfrag}
\usepackage{bm}
\usepackage{epstopdf}
\usepackage{latexsym}
\usepackage{multirow}
\usepackage{amsmath,amssymb,amsfonts}
\usepackage{enumerate}
\usepackage[utf8]{inputenc}
\usepackage{color,soul}
\usepackage{float}
\usepackage[utf8]{inputenc}
\usepackage[utf8]{inputenc}
\usepackage[utf8]{inputenc}
\usepackage{tikz}
\usepackage[utf8]{inputenc}
\begin{document}
\date{\today}

\title{ Stability of patterns on thick curved surfaces}
\author{Sankaran N.
\footnote{sankaran@iisertvm.ac.in}}
\affiliation{ School of Physics, Indian Institute of Science Education and Research, Thiruvananthapuram-695016, India}
\begin{abstract}
We consider reaction-diffusion equations on a thick curved surface and obtain
a set of effective R-D equation to ${\cal O}(\epsilon^2)$,
where $\epsilon$ is the surface thickness. We observe that the
R-D systems on these curved surfaces can have space-
dependent reaction kinetics. Further, we use linear stability analysis to study
the Schnakenberg model on spherical
and cylindrical geometries. The  dependence of
steady state on the thickness is determined for both cases, and
we find that a change in the thickness can 
stabilize the unstable patterns, and vice versa. The combined effect
of thickness and curvature
can play an important role in the rearrangement of spatial patterns on thick curved
surfaces.
 
\end{abstract}
\maketitle
\section{INTRODUCTION}
\paragraph*{}
In 1952 Turing~\cite{turing} proposed the reaction-diffusion\\(R-D) mechanism,
where the chemicals can react and diffuse so as to produce
spatially varying concentrations of chemicals in the steady state. Since then 
many models~\cite{murray,biology} have been proposed to mimic the complex pattern formation
in biological systems. Turing-like R-D equations are routinely used in trying to
understand the skin patterns of
animals~\cite{murray}; for example 
  in fish~\cite{shoji}, mammals~\cite{bard}, snakes~\cite{snake} leopards~\cite{maini} and many others.
There have also been attempts to study changes in the pigmentation
patterns on leopards and jaguars as they grow in size~\cite{maini}.
\paragraph*{}
In most studies, R-D equations are analyzed on flat geometries
which are not always suitable for the study of patterns on animal skin surfaces. 
It is reasonable to assume that geometry of the surface
can play an important role in determining the pattern formation.
For instance, Turing considered the surface of a sphere 
in the context of gastrulation of a blastula~\cite{turing}. 
Geometry is probably responsible for stripes at the end of the tail
while there are spots elsewhere in some animals~\cite{murray}. Understanding
the pattern formation 
on curved surfaces can be important in some
 chemical, biochemical and embryological process~\cite{varea}. It is also known that 
organ morphogenesis can be controlled
 by tissue geometry~\cite{nelson}. Recently some studies have been initiated to
 understand the role of curvature in biological systems~\cite{hal,peter,gov,orland}.
\paragraph*{}
The thickness of the surface is being neglected in most earlier studies. For instance, 
blastula is considered as a hollow sphere 
assuming no thickness~\cite{turing}. But protein diffusion in lipid bilayers
can be viewed as a diffusion on a two-dimensional curved surface with 
thickness.
In some of the recent studies,
 there has been attempts to incorporate the small
 thickness~\cite{ogawa_curvature-dependent_2010,balois}. The combined
 role of geometry and thickness can lead to curvature-dependent
 diffusion
 and may result in complex pattern formation in animals~\cite{ogawa_curvature-dependent_2010}. 
 Recent study on
 the pattern formation in melanocytic tumours again suggests the importance of geometry
 and of the thickness~\cite{balois}.
 \paragraph*{}
   Some of these studies suggest that it is relevant to ask about the 
   effect of the thickness and curvature in R-D systems. In particular,
    how does the curvature and thickness affect the formation of steady state patterns?.
   We answer this question by considering the reaction-diffusion
   of two chemicals on a curved surface with small thickness. 
   We first obtain an effective description of
   R-D equation and then 
   deduce the dependence of steady state and its stability on the thickness.
 \paragraph*{}
 This paper is organized as follows. In sec. II, we describe a general model
 of a R-D system and then explicitly obtained its effective description.
  In sec. III, we analyze the effect of the thickness and curvature in the
  Schnakenberg model, specifically on 
    a spherical and cylindrical geometry. Finally, 
 we summarize our results in sec. IV.
\section{ model}
We consider reaction-diffusion of two chemicals 
between two  curved surfaces, $\sigma$ and $\sigma^{'}$, which are
parallel to each other and separated by a distance $\epsilon$. The concentrations of chemicals
are denoted by $A(q^{0},q^{1},q^{2},t)$ and $B(q^{0},q^{1},q^{2},t)$ and the 
dynamics is governed by the R-D equations~\cite{turing}
\begin{eqnarray}
 \frac{\partial A}{\partial t}=F_{1}(A,B)+D_{A}\nabla^{2} A\\
 \frac{\partial B}{\partial t}=F_{2}(A,B)+D_{B}\nabla^{2} B
\end{eqnarray}
where $F_{1}(A,B)$ and $F_{2}(A,B)$ are the the reaction kinetics, which in general  are nonlinear functions; $D_{A}$ and $D_{B}$ are the
diffusion constants of the chemicals.
\begin{figure}[h]
 \centering
\includegraphics[height=30mm,width=35mm]{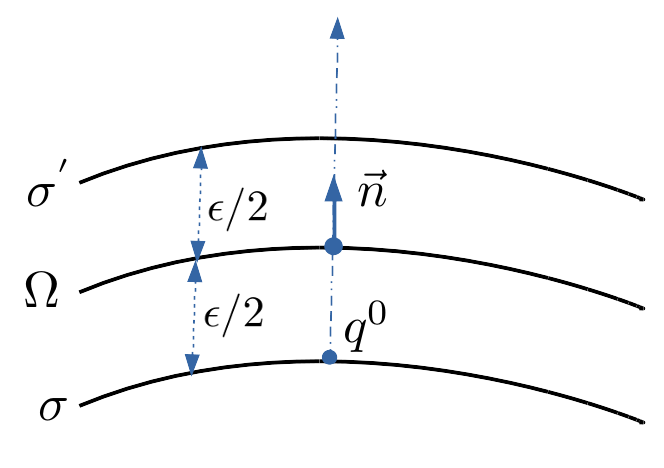}
\caption{ Coordinate System}
\end{figure}
\paragraph*{}
The co-ordinate system we use here is similar to the one considered by Ogawa~\cite{ogawa_curvature-dependent_2010}.
We place a curved surface $\Omega$ between the surfaces $\sigma^{'}$ and $\sigma$ at distances
$\epsilon/2$ and $-\epsilon/2$ respectively. Any point between the surfaces
$\sigma$ and $\sigma^{'}$ can be represented by $q^{\mu}=(q^{0},q^{1},q^{2})$
where $(q^{1},q^{2})$ are the curved co-ordinates on the surface $\Omega$ and
$q^{0}$ is the normal co-ordinate. The components of the metric $G_{\mu\nu}$ of the curvilinear 
co-ordinate system can be given as

\begin{eqnarray}
G_{0i}=G_{i0}=0, ~~G_{00}=1,\nonumber\\
G_{ij}=g_{ij} + 2 q^0 \kappa_{ij} + (q^0)^2 \kappa_{im} \kappa^m_j,
\end{eqnarray}
where $g_{ij}$ is the metric on the curved surface $\Omega$, $\kappa_{ij}$
is the second fundamental tensor.
The determinant of the metric, $G=\det(G_{\mu\nu}$) can be written as,
\begin{equation}
 G_{~~} = g~ \{1 + 2 q^0 \kappa + (q^0)^2 (\kappa^2 + R) + {\cal O}(\epsilon^3)\},
\end{equation}
where, $g=\det(g_{ij})$, mean curvature $\kappa=g^{ij} \kappa_{ij}$ and 
Ricci scalar $R=(\kappa^2 -\kappa_{ij} \kappa^{ij})$.

\subsection{ Effective Theory}
\paragraph*{}
In this subsection, we obtain the effective two dimensional description of R-D equations.
\paragraph*{}
 The total amount of the chemicals  present in the system can be decomposed as follows
\begin{eqnarray}
  \int A  \sqrt{G}~ d^3 q
= \int [\int_{-\epsilon/2}^{\epsilon/2} d q^0  A~ \sqrt{\frac{G}{g}}~] ~ \sqrt{g} ~ d^2 q ,\\
\int B  \sqrt{G}~ d^3 q
= \int [\int_{-\epsilon/2}^{\epsilon/2} d q^0  B~ \sqrt{\frac{G}{g}}~] ~ \sqrt{g} ~ d^2 q,
\end{eqnarray}
thus leading to a definition of concentrations of chemicals in the effective description
\begin{eqnarray}
 \tilde{A}(q^1,q^2,t)=\int_{-\epsilon/2}^{\epsilon/2} d q^0 ~ \sqrt{\frac{G}{g}}~ A(q^{0},q^{1},q^{2},t),\\
 \tilde{B}(q^1,q^2,t)=\int_{-\epsilon/2}^{\epsilon/2} d q^0 ~ \sqrt{\frac{G}{g}}~ B(q^{0},q^{1},q^{2},t),
\end{eqnarray}
Multiplying equations (1) and (2) with $\sqrt{G/g}$ and integrating over $q^{0}$ 
result in the equations
\begin{eqnarray}
 \frac{\partial \tilde{A}}{\partial t}=\tilde{F_{1}}(\tilde{A},\tilde{B})+D_{A}\nabla^{2}_{eff} \tilde{A},\\
 \frac{\partial \tilde{B}}{\partial t}=\tilde{F_{2}}(\tilde{A},\tilde{B})+D_{B}\nabla^{2}_{eff}\tilde{ B},
\end{eqnarray}
where
\begin{eqnarray}
 \tilde{F_{1}}(\tilde{A},\tilde{B})=\int_{-\epsilon/2}^{\epsilon/2} d q^0 ~ \sqrt{\frac{G}{g}}~ F(A,B),\\
 \tilde{F_{2}}(\tilde{A},\tilde{B})=\int_{-\epsilon/2}^{\epsilon/2} d q^0~ \sqrt{\frac{G}{g}} ~G(A,B),
\end{eqnarray}
and 
\begin{eqnarray}
 \nabla_{eff}^{2}\tilde{A}=\int_{-\epsilon/2}^{\epsilon/2} d q^0~ \sqrt{\frac{G}{g}}~\nabla^{2}A,\\
 \nabla_{eff}^{2}\tilde{B}=\int_{-\epsilon/2}^{\epsilon/2} d q^0 ~\sqrt{\frac{G}{g}}~\nabla^{2}B,
\end{eqnarray}
If we assume that concentrations of chemicals are independent of
$q^{0}$ co-ordinate, then we obtain to ${\cal O}(\epsilon^2)$

\begin{eqnarray}
 \tilde{A}=\epsilon(1+\frac{\epsilon^{2}}{24}R)~A(q^{1},q^{2}),\\
 \tilde{B}=\epsilon(1+\frac{\epsilon^{2}}{24}R)~B(q^{1},q^{2}),
\end{eqnarray}
 and hence  equations (11) and (12) can be rewritten as
\begin{flushleft}
\begin{eqnarray}
 \tilde{F_{1}}(\tilde{A},\tilde{B})\!\!\!&=&\!\!\!\epsilon(1\!\!+\!\!\frac{\epsilon^{2}}{24}R)
 F_{1}(\frac{1}{\epsilon}(1\!-\!\frac{\epsilon^{2}}{24}R)\tilde{A},\frac{1}{\epsilon}(\!1-\!\frac{\epsilon^{2}}{24}R)\tilde{B}),\nonumber\\
 \\
 \tilde{F_{2}}(\tilde{A},\tilde{B})\!\!\!&=&\!\!\! \epsilon(1\!+\!\frac{\epsilon^{2}}{24}R)
 F_{2}(\frac{1}{\epsilon}(1\!-\!\frac{\epsilon^{2}}{24}R)\tilde{A},\frac{1}{\epsilon}(1\!-\!\frac{\epsilon^{2}}{24}R)\tilde{B}).\nonumber
 \\
\end{eqnarray}
\end{flushleft}

Assuming the fluxes $\nabla A$ and $\nabla B$ vanish at
boundaries $\sigma$ and $\sigma^{'}$, and following 
similar steps as in~\cite{ogawa_curvature-dependent_2010} for 
the equations (13) and (14) will lead to
\begin{flushleft}
 \begin{eqnarray}
  \nabla_{eff}^{2}&=&\Delta^{(2)}   + \tilde{\nabla}\\
  \tilde{\nabla}&=&\frac{\epsilon^{2}}{12} g^{-1/2} \frac{\partial}{\partial q^i} ~ g^{1/2}\nonumber\\
 &\times& \{ (3 \kappa^{im} \kappa_m^j -2 \kappa \kappa^{ij}) \frac{\partial}{\partial q^j} 
- \frac{1}{2} g^{ij} \frac{\partial R}{\partial q^j} \}\nonumber.
 \end{eqnarray}

\end{flushleft}
where $\Delta^{(2)} $ is the Laplace-Beltrami operator on the curved surface $\Omega$. To
summarize, the effective description is captured in (9),(10),(17),(18) and (19). 

\section{Effect of thickness and curvature}
\paragraph*{}
In this section, we illustrate the effect of the thickness and
curvature in the effective description of a R-D equation. 
In particular, how does the nature of steady state
vary by changing the thickness?. This question can be addressed within 
the framework of above discussed effective theory.
Note that, the thickness is kept
constant during the dynamics of the system.
\paragraph*{}
In flat geometries, it can be seen from equations (15) and (16), both the concentration
are scaled, $A\rightarrow \epsilon A$ and $B\rightarrow \epsilon B$,
and dynamics remains the same. While in curved geometries, 
combined role of thickness and curvature can lead to nontrivial
effects.
\paragraph*{}
In the following, we consider a R-D system between two curved surfaces, and
point out some salient features of the effective description. We then proceed to
analyze the effective description on spherical, and
cylindrical geometry.
\subsection{General surface}
Let us restrict to the well-studied Schnakenberg model, where the reaction kinetics
is taken as~\cite{schnakenberg}
\begin{eqnarray}
F_{1}(A,B) =k_{1}-k_{2}A+k_{3}A^{2}B,\\
 F_{2}(A,B)=k_{4}-k_{3}A^{2}B,
\end{eqnarray}
\paragraph*{}
Here, we confine the chemicals between two curved surfaces $\sigma$ and 
$\sigma^{'}$ placed at a distance $\frac{\epsilon}{2}$ from a general curved
surface $\Omega$.
We can read the effective description of Schnakenberg model on a general surface
using equations (17) and (18), and is given by
\begin{eqnarray}
 \frac{\partial \tilde{A}}{\partial t}&=&\tilde{k}_{1}
 -\tilde{k}_{2}\tilde{A}+\tilde{k}_{3}\tilde{A}^{2}\tilde{B}
 +D_{A}\nabla^{2}_{eff}\tilde{A},\\
 \frac{\partial \tilde{B}}{\partial t}&=&\tilde{k}_{4}
 -\tilde{k}_{3}\tilde{A}^{2}\tilde{B}+D_{B}\nabla^{2}_{eff}\tilde{B},
\end{eqnarray}
where $\nabla^{2}_{eff}$ is given by (19), and the reaction constants in effective theory 
are related to the original model as
\begin{eqnarray*}
 \tilde{k}_{1}&=&\epsilon k_{1}(1+\frac{\epsilon^{2}}{24}R),~
 \tilde{k}_{3}=\frac{1}{\epsilon^{2}}k_{3}(1-\frac{\epsilon^{2}}{12}R),~\\
 \tilde{k}_{4}&=&\epsilon k_{4}(1+\frac{\epsilon^{2}}{24}R),~
 \tilde{k}_{2}=k_{2}~.
\end{eqnarray*}
\paragraph*{}
 In general, the reaction constants in an effective theory of the Schnakenberg model
are space-dependent as the Ricci scalar$(R)$ is not necessarily a constant,
and may lead to interesting consequences. The 
space-dependent reaction kinetics can result in an absence of a homogeneous
steady state.
  
 
\paragraph*{} 
 A few comments about the dependence of reaction rates on Ricci scalar follows. 
The term $\tilde{k}_{3}\tilde{A}^{2}\tilde{B}$ in equation (22) represents the production of the 
chemical $\tilde{A}$. Note that the reaction constant $ \tilde{k}_{3}$ is lower
in regions with higher positive curvature.
Hence, the production of $\tilde{A}$ is more in regions with lower positive
curvature. Similarly
the term -$\tilde{k}_{3}\tilde{A}^{2}\tilde{B}$ in equation (23) represents 
the depletion of $\tilde{B}$ and hence can result in more
depletion in regions of lower positive curvature. Both the reaction constants $\tilde{k}_{1}$ and $\tilde{k}_{4}$ 
are higher in regions with higher positive curvature and result in 
 more production of chemicals in these regions. 

\paragraph*{}
The term $D_{A}\nabla_{eff}^{2}$ can be rewritten as 
\begin{equation} 
 \frac{1}{\sqrt{g}}\partial_{i}\sqrt{g}(D_{\tilde{A}}^{ij}\partial_{j}\tilde{A}
 -\frac{\epsilon^{2}}{24} g^{ij}\partial_{j}R.\tilde{A}),
\end{equation}
where  
\begin{equation}
 D_{\tilde{A}}^{ij}=D_{A}(g^{ij}+\frac{\epsilon^{2}}{12}\{3\kappa^{im}\kappa_{m}^{j}
 -2\kappa\kappa^{ij}\}).
\end{equation}
\paragraph*{}
The first term in equation (24) is the diffusion term,
 which is not necessarily isotropic, and is characterized by the diffusion
 matrix $D^{ij}$ which depends on the extrinsic curvature.
  On a sphere the ${\cal O}(\epsilon^2)$ term of both $D^{\theta\theta}$ and
  $D^{\phi\phi}$ is negative, 
  where $(\theta,\phi)$ are the co-ordinates on the surface of a sphere.
  But on a cylindrical surface the ${\cal O}(\epsilon^2)$ term
  of $D^{\theta\theta}$ is positive and $D^{zz}=0$,
  where $(\theta,z)$ are the co-ordinates on the surface of a cylinder. 
  Hence there is
an enhanced
diffusion of chemicals along the the $\theta$ direction on a locally cylindrical region.
The diffusion of chemical can be slow down along $\theta$ and $\phi$ 
directions on locally spherical regions. 
 The second term in the equation (24) is the current due to the gradient of
 Ricci scalar between two points on a surface~\cite{ogawa_curvature-dependent_2010}.
\subsection{ Spherical Geometry}
 We now consider the Schnakenberg model, where the 
chemicals are confined to the region between two spheres of radii 
 $a_{0}+\frac{\epsilon}{2}$ and $ a_{0}-\frac{\epsilon}{2}$. Choose
 the  sphere with radius $a_{0}$ as the surface $\Omega$
 and spheres with radii  $a_{0}+\frac{\epsilon}{2}$ and $~$ $ a_{0}-\frac{\epsilon}{2}$ 
 are the surfaces $\sigma^{'}$ and $\sigma$, respectively.
 Our analysis suggests the following effective description for this
 model

\begin{eqnarray}
 \frac{\partial \tilde{A}}{\partial t}=\tilde{k_{1}}
 -\tilde{k_{2}}\tilde{A}+\tilde{k_{3}}\tilde{A}^{2}\tilde{B}
 +\nabla^{2}_{eff}\tilde{A},\\
 \frac{\partial \tilde{B}}{\partial t}=\tilde{k_{4}}
 -\tilde{k_{3}}\tilde{A}^{2}\tilde{B}+\nabla^{2}_{eff}\tilde{B},
\end{eqnarray}
where the reaction constants in the effective theory are related to original model
as follows
\begin{eqnarray*}
 \tilde{k_{1}}&=&k_{1}\epsilon(1+\frac{\epsilon^{2}}{24a_{0}^{2}}),~
 \tilde{k_{3}}=\frac{1}{\epsilon^{2}}k_{3}(1-\frac{\epsilon^{2}}{12a_{0}^{2}}),~\\
 \tilde{k_{4}}&=&k_{4}\epsilon(1+\frac{\epsilon^{2}}{24a_{0}^{2}}),~
 \tilde{k_{2}}=k_{2}~,
\end{eqnarray*}
since $R=1/a_{0}^{2}$ for the sphere of radius $a_{0}$.\\
Choosing the $(\theta,\phi)$ coordinates  on the surface of a sphere with 
radius $a_{0}$, it is straightforward to obtain
\begin{eqnarray*}
 g_{\theta\theta}=a_{0}^{2}, g_{\phi\phi}=a_{0}^{2}\sin^{2}\theta, g_{\theta\phi}=g_{\phi\theta}=0,\\
\kappa_{\theta\theta}=-a_{0}, \kappa_{\phi\phi}=-a_{0} \sin^{2}\theta, \kappa_{\theta\phi}=\kappa_{\phi\theta}=0,\\
\kappa_{\phi}^{\phi}=\frac{-1}{a_{0}}, \kappa_{\theta}^{\theta}=\frac{-1}{a_{0}}, \kappa_{\theta}^{\phi}=\kappa_{\phi}^{\theta}=0.
\end{eqnarray*}
Hence $\nabla^{2}_{eff}$ read as
\begin{equation}
 \nabla^{2}_{eff}=\frac{1}{b_{0}^{2}}\{\frac{\partial^{2}}{\partial \phi^{2}}
 +\frac{1}{\sin\theta}\frac{\partial}{\partial \theta}(\sin\theta ~\frac{\partial}{\partial
 \theta})\},
\end{equation}
where $b_{0}=a_{0}(1+\frac{\epsilon^{2}}{24 a_{0}^{2}})$. In essence, the effective 
two dimensional R-D equations (26) and (27) can be interpreted as
Schnakenberg model on a sphere of radius $b_{0}$ with redefined parameters.

 Equations (26) and (27) can be rewritten in terms of rescaled variables as

\begin{eqnarray}
 \frac{\partial U}{\partial \tau}=\gamma(a-U+
 U^{2}~V)+\tilde{\nabla}_{eff}^{2}U,\\
 \frac{\partial V}{\partial \tau}=\gamma(b-U^{2}~V)
 +d~\tilde{\nabla}_{eff}^{2}{V},
\end{eqnarray}
where,
\begin{eqnarray}
 \tau=D_{A}t/b_{0}^{2},U=\tilde{A}(\frac{\tilde{k}_{3}}{\tilde{k}_{2}})^{1/2},~
 V=\tilde{B}(\frac{\tilde{k}_{3}}{\tilde{k}_{2}})^{1/2},~
 d=\frac{D_{B}}{D_{A}},\nonumber\\
 \\
 a=(\frac{\tilde{\kappa}_{1}}{\tilde{k}_{2}})(\frac{\tilde{k}_{3}}{\tilde{k}_{2}})^{1/2}
 ,~b=(\frac{\tilde{k}_{4}}{\tilde{k}_{2}})(\frac{\tilde{\kappa}_{3}}{\tilde{k}_{2}})^{1/2}
 ,~\gamma=\frac{b_{0}^{2}\tilde{\kappa_{2}}}{D_{A}}.\nonumber
\end{eqnarray}
 and $\tilde{\nabla}_{eff}^{2}$ is the Laplace operator in the scaled variables.


The homogeneous steady state solution can be obtained from (29) and (30) as
$(U_{0},V_{0})=(a+b,\frac{b}{(a+b)^{2}})$. We consider only positive solution
of homogeneous steady state and discard the solution with negative concentration. Note that homogeneous steady state solution is independent of the thickness to ${\cal O}(\epsilon^2)$. 
\paragraph*{}
The linear stability analysis about the homogeneous steady state follows.
A small variation in the homogeneous steady state is denoted as
\[\delta W=\left (\begin{array}{c}
         \delta U-U_{0}\\
         \delta V-V_{0}\\
         \end{array}\right),\]
         
         which satisfies the linearized equation
         \begin{equation}
          \frac{\partial~ (\delta W)}{\partial t}=\hat{L}\delta W,
         \end{equation}
         where
         \begin{equation}
          \hat{L}=\gamma C+D\nabla_{eff}^{2},
         \end{equation}
\[D=\left (\begin{array}{cc}                      
         1&0\\
         0&d\\
         \end{array}\right),
         C=\left (\begin{array}{cc}
         \frac{\partial f}{\partial U}&\frac{\partial f}{\partial V}\\
         \\
         \frac{\partial g}{\partial U}&\frac{\partial g}{\partial V}\\
         \end{array}\right)_{U_{0},V_{0}},\]


and
\begin{eqnarray*}
 f(U,V)=\gamma(a-U+
 U^{2}~V),\\
 g(U,V)=\gamma(b-U^{2}~V)~.
\end{eqnarray*}
         \paragraph*{}
         The solution to the equation (32) can be written as\\
         \begin{equation}
          \delta W(\theta,\phi,t)=\sum_{l=0}^{l=\infty}\sum_{m=-l}^{l}C_{l}^{m}e^{\lambda(l) t}
          P_{l}^{m}(\cos\theta)e^{im\phi},
         \end{equation}
where the constants $C_{l}^{m}$ can be determined from initial conditions. The
eigenvalues $\lambda(l)$ satisfy
\begin{equation}
 \lambda^{2}+\lambda[(l(l+1))(1+d)-\gamma(f_{u}+g_{v})]+h(l(l+1))=0,
\end{equation}
where $f_{u}=\frac{\partial f}{\partial U}$, $g_{v}=\frac{\partial g}{\partial V}$
and $h(l(l+1))$ can be given as
\begin{equation}
 h(l(l+1))=d(l(l+1))^{2}-\gamma(d~f_{u}+g_{v}) ~l(l+1)+\gamma^{2}(f_{u}g_{v}-f_{v}g_{u}).
\end{equation}

The necessary condition for the instability to kick in is
\begin{equation*}
 h(l(l+1))<0,
\end{equation*}


\paragraph*{}
The modes $l$ which satisfy $h(l(l+1))<0$ gives positive eigenvalues in equation (35).
These are the modes which give rise to the instability (an inhomogeneous steady state) to the system.
These modes(unstable modes) lie between $L_{-}<l(l+1)<L_{+}$, where $L_{-}$ and $L_{+}$ are the
roots of $h(l(l+1))=0$, and given as
\begin{eqnarray}
 L_{\pm}(\epsilon)=\frac{a_{0}^{2}\tilde{k_{2}}}{D_{A}.2d}(1+\frac{\epsilon^{2}}{12a_{0}^{2}})[(df_{u}+g_{v})\nonumber\\\pm\{(df_{u}+g_{v})^{2}-4d(f_{u}g_{v}-f_{v}g_{u})\}^{1/2}].
\end{eqnarray}
\begin{figure}[H]
\centering
\includegraphics{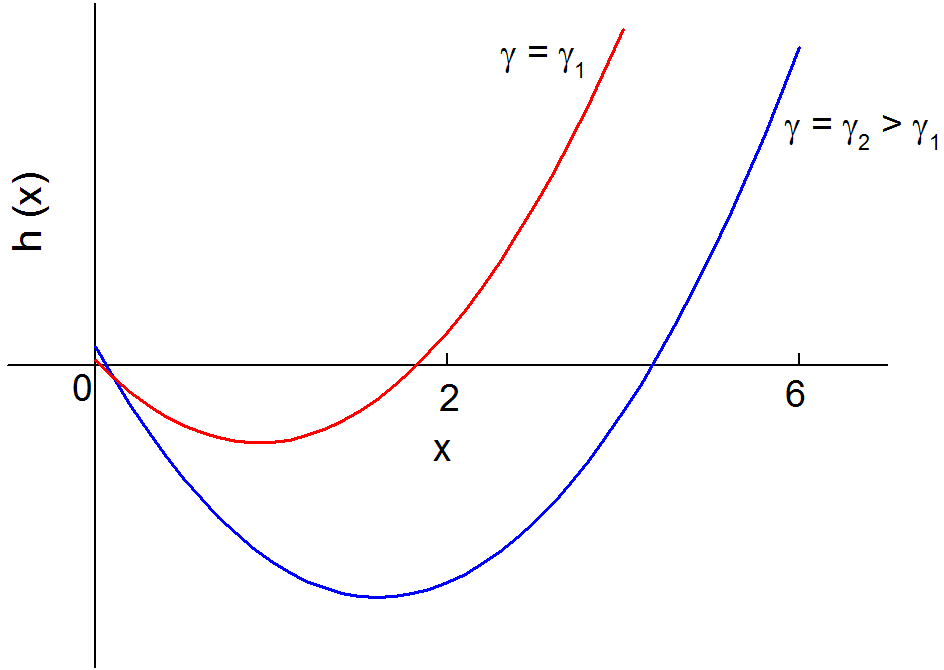}
\caption{h(x) Vs x: Figure illustrates the role of thickness-dependent $\gamma$ in the selection
of unstable modes.}
\end{figure}
\paragraph*{}
 Note that the unstable modes lie between the zeros$(L_{+},L_{-})$ of the
function $h(x)$, where $L_{+}$ and $L_{-}$ depend on the thickness.
It can be seen from the equation (37) that an increase(decrease) in 
thickness can shift the zeros of $h(x)$ towards right(left) on 
the $x$ axis as shown in the fig(2), and this can result in a scenario
where the stable modes can become unstable and vice versa. Hence the 
 nature of steady state can be changed by tuning
the thickness. In other words, there can be
a transition from a homogeneous steady state to an inhomogeneous steady state(a pattern)
by changing the thickness.
It is also possible to obtain different
inhomogeneous steady states as thickness changes. The $\epsilon$ dependence in the equation
(37) 
thus reveals the effect
of thickness on the stability.

\paragraph*{}
Let us distinguish two cases. In the case I, for 
certain ranges of parameters(a,b,d,$\gamma$)
the zeros of function $h(x)$ lie below the point $x=2$,
 namely, $0< L_{-}$ and $L_{+} <2$.
Here, the steady state configuration is homogeneous.
 In this case, since there
is no unstable mode lies below the point $x=2$,  the instability
can set in only by increasing the thickness.
\paragraph*{}
In the case II, the zeros of function $h(x)$ lie between
 $l_{1}(l_{1}+1)$ and $(l_{1}+1)(l_{1}+2)$, namely,
 $l_{1}(l_{1}+1)<L_{-}$ and $L_{+}<(l_{1}+1)(l_{1}+2)$. In this case the 
system can be driven to an inhomogeneous steady state either by decreasing
or increasing the thickness.
\subsection{ Cylindrical Geometry}
Here we consider Schnakenberg model, where 
the chemicals are confined between two cylinders of radii $a_{0}+\frac{\epsilon}{2}$ and
$a_{0}-\frac{\epsilon}{2}$, 
and further assume the flux vanishes at $z=0$ and $z=l$.

 In this case the effective description is governed by
 \begin{eqnarray}
 \frac{\partial \tilde{A}}{\partial t}&=&\tilde{\kappa}_{1}
 -\tilde{\kappa}_{2}\tilde{A}+\tilde{\kappa}_{3}\tilde{A}^{2}\tilde{B}
 +\nabla^{2}_{eff}\tilde{A},\\
 \frac{\partial \tilde{B}}{\partial t}&=&\tilde{\kappa}_{4}
 -\tilde{\kappa}_{3}\tilde{A}^{2}\tilde{B}+\nabla^{2}_{eff}\tilde{B},
\end{eqnarray}
where the reaction constants in the effective theory are related to the original
model as follows
\begin{eqnarray*}
 \tilde{k}_{1}=k_{1}\epsilon,~
 \tilde{k}_{3}=\frac{1}{\epsilon^{2}}k_{3},~
 \tilde{k}_{4}=k_{4}\epsilon,~
 \tilde{k}_{2}=k_{2}~.
\end{eqnarray*}
 on the surface of a cylinder with radius $a_{0}$ the quantities related to
 intrinsic and extrinsic curvatures are
\begin{eqnarray*}
 g_{\theta\theta}=a_{0}^{2},g_{zz}=1,g_{\theta z}=g_{z \theta}=0,\\
 \kappa_{\theta\theta}=a_{0},\kappa_{zz}=0,\kappa_{\theta z}=\kappa_{z \theta}=0,\\
 \kappa^{\theta\theta}=\frac{1}{a_{0}^{3}},\kappa^{zz}=0,\kappa^{\theta z}=0.
\end{eqnarray*}
Hence,
\begin{equation}
 \nabla^{2}_{eff}=\frac{1}{b_{0}^{2}}\frac{\partial^{2}}{\partial \theta^{2}}+
 \frac{\partial^{2}}{\partial z^{2}},
\end{equation}
where $b_{0}=a_{0}\{1-\frac{\epsilon^{2}}{24 a_{0}^{2}}\}$. Thus the
effective equations $~$(38) and (39) can be interpreted as 
Schnakenberg model on a cylinder
with rescaled radius $b_{0}$.
Equations (38) and (39) can be rewritten in terms of rescaled variables, 
$U=U(\theta,\tilde{z})$, and $V=V(\theta,\tilde{z})$, obeying the equations
\begin{eqnarray}
 \frac{\partial U}{\partial \tau}=\gamma(a-U+
 U^{2}~V)+\tilde{\nabla}_{eff}^{2}U,\\
 \frac{\partial V}{\partial \tau}=\gamma(b-U^{2}~V)
 +d~\tilde{\nabla}_{eff}^{2}{V}~,
\end{eqnarray}
where the scaled variables are defined in equation~(31) and $\tilde{z}=\frac{z}{b_{0}}$, and
$\tilde{\nabla}_{eff}^{2}$ is Laplace operator in the scaled variables.
 \paragraph*{}
 If we now proceed similar to the case of a sphere, then
 the deviation from homogeneous solution
         \begin{equation}
          \delta W(\theta,\tilde{z},t)=
          \sum_{n,m}C_{nm}e^{\lambda t}e^{i n\theta }\cos(\frac{m\pi b_{0}}{l}\tilde{z}),
         \end{equation}
         where $C_{nm} $ depends on the initial conditions. The eigenvalues 
         can be obtained from
 \begin{equation*}
 \lambda^{2}+\lambda[(n^{2}+\frac{m^{2}\pi^{2}b_{0}^{2}}{l^{2}})(1+d)-\gamma(f_{u}+g_{v})]+h(n,m)=0,
\end{equation*}
where
\begin{eqnarray*}
 h(n,m)=d\{n^{2}+\frac{m^{2}\pi^{2}b_{0}^{2}}{l^{2}}\}^{2}-\gamma(d~f_{u}+g_{v})(n^{2}+\frac{m^{2}\pi^{2}b_{0}^{2}}{l^{2}}) \\~+\gamma^{2}(f_{u}g_{v}-f_{v}g_{u}).
\end{eqnarray*}\\
Following the same analysis as done in the case of a sphere, the modes (n,m) 
which satisfy the following condition
\begin{equation}
 R_{-}^{2}<(n^{2}+\frac{m^{2}\pi^{2}b_{0}^{2}}{l^{2}})<R_{+}^{2},
\end{equation}
will destabilize the homogeneous solution,
where $R_{-}^{2}$ and $R_{+}^{2}$ are the zeros of $h(n,m)$ and given by
\begin{eqnarray}
 R_{\pm}^{2}(\epsilon)=\frac{a_{0}^{2}\tilde{k_{2}}}{D_{A}.2d}(1-\frac{\epsilon^{2}}{12a_{0}^{2}})
 [(df_{u}+g_{v})\nonumber\\\pm\{(df_{u}+g_{v})^{2}-4d(f_{u}g_{v}-f_{v}g_{u})\}^{1/2}]~.
 \end{eqnarray}
Hence the unstable modes $\displaystyle(n,m)$
 lie between the two semicircles 
 in the $\displaystyle(n,\frac{m\pi b_{0}}{l})$ plane  
  with radius $R_{+}$ and $R_{-}$ as shown in the figure 3. 
  It can be seen from equation (45) that the values of both $R_{-}$ and $R_{+}$ 
 decrease by increasing the thickness($\epsilon$).
 \begin{figure}[H]
\centering
\includegraphics[height=50mm,width=50mm]{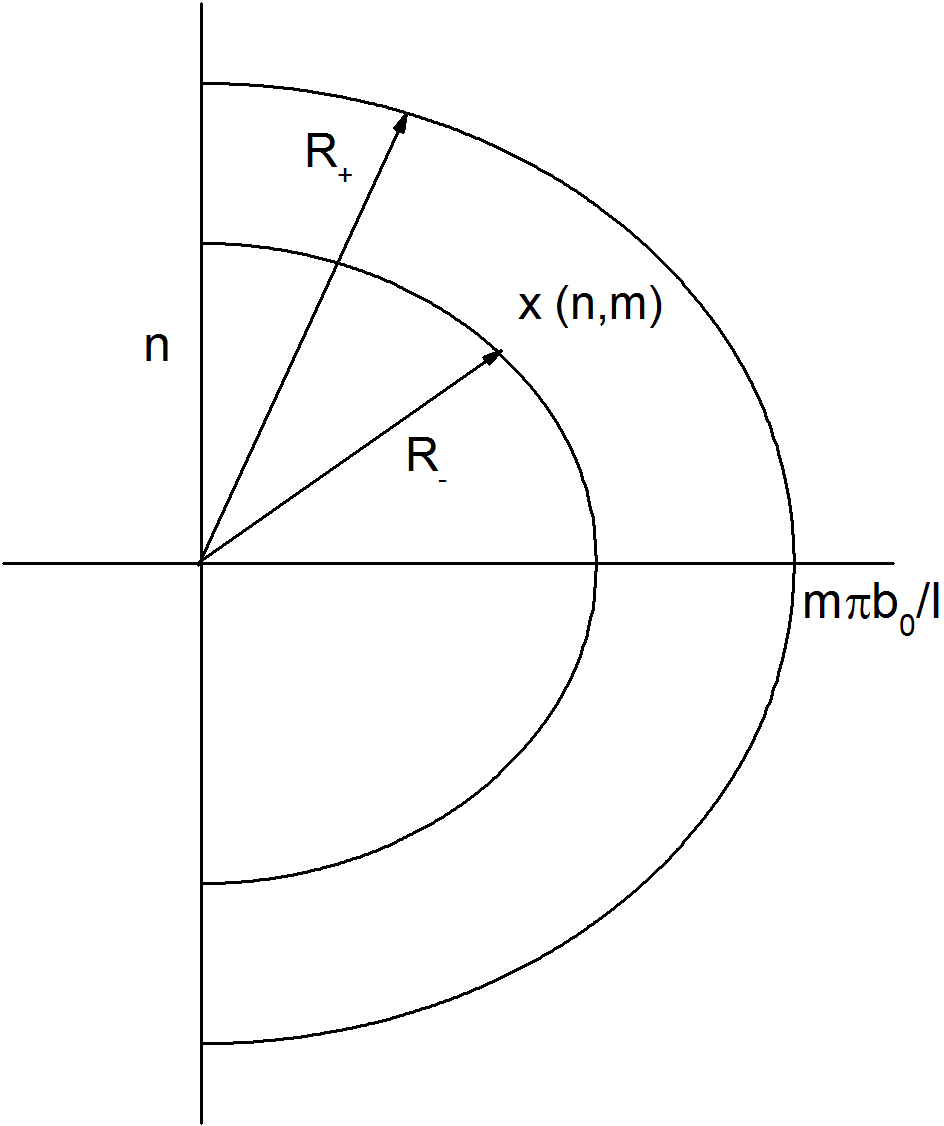}
\caption{Unstable modes $(n,m)$ lie between semicircles with radius
$R_{-}$ and $R_{+}$. }
\end{figure}

\paragraph*{}
Let us distinguish two cases. 
In the case I, 
the space between two semicircles in the $(n,\frac{m \pi b_{0}}{l})$ plane
encloses no mode $(n,m)$ satisfying the condition (44). The steady state 
configuration of R-D system is homogeneous in this case. By increasing 
the thickness the semicircles can shrink, and can result in the inclusion of some
modes $(n,m)$. Hence, there is a possibility
of obtaining an inhomogeneous steady state from the homogeneous one 
by increasing the thickness. It is also possible in this case
that the region between two circles can encompass no modes $(n,m)$ even
after increasing the thickness. In such a situation, the system
can continue to be in the homogeneous steady state.
\paragraph*{}
In the case II, some modes $(n,m)$ are enclosed
within the semicircles of radius $R_{+}$ and $R_{-}$. The steady state configuration
is inhomogeneous in this case. Now an increase in thickness can result in the shifting of curves 
such that modes $(n,m)$ are no longer present 
between $R_{-}$ and $R_{+}$ semicircles. Hence, there is a possibility
of transition to a homogeneous steady state from an inhomogeneous one
by increasing the thickness. Another possibility is that the semicircles can 
shrink.
This can lead to an inclusion of new modes $(n,m)$ and result in transition to
 a new inhomogeneous steady state
 from the initial inhomogeneous state.


\section{Conclusion}
To conclude, we have studied the effect of curvature and thickness
 in R-D systems on quasi-two dimensional(thick) curved surfaces.
 We explicitly analyzed an effective description of
the Schnakenberg model, in particular, on  spherical and 
cylindrical geometry. In both spherical and cylindrical case, the effective theory 
is same as the original model
on a sphere and a cylinder, respectively, with rescaled parameters.
 On the spherical geometry an increase in the thickness can lead to an increase
 in the parameter $\gamma$. In cylindrical geometry the parameter $\gamma$
 can decrease by increasing the thickness. In the absence of curvature, the thickness
 play no significant role in the effective description.
\paragraph*{}
 
In
general,  R-D systems on quasi-two dimensional curved surfaces
can have space-dependent parameters.
 There are a few instances where spatially varying parameters
 are considered~\cite{complex,bhatta,page,benson,spatial}. The 
absence of homogeneous steady state is also a characteristic of the effective R-D equation.
 The effective R-D description is not necessarily 
 similar to the original description on a two-dimensional surface
 with rescaled parameters.
 In R-D systems on quasi-two 
 dimensional curved surfaces, a change in thickness can stabilize
 the unstable patterns and vice versa. Hence the patterns
 (inhomogeneous steady state) can appear or disappear by tuning the thickness.
 This might be a plausible reason for different patterns on leopards and jaguars~\cite{maini} as they grow
 in size or rather as the skin thickness increases.
\paragraph*{}
There is a related model studied in the context of
Belousov-Zhabotinsky reaction~\cite{r} where there is no diffusion. Instead,
equation to the chemical $A$ contain 
$v.\nabla A$  term, where $v$ is the velocity of the chemical $A$,
while the other chemical $B$ is immobilized. In this model the chemical instability
is of traveling-wave type, and the concentrations can vary both in space and time. In this case
there 
is an isotropy in the wave speed when the speed 
of the chemical $A$ is same in every directions. Assume
 that the velocity$(v)$ of the chemical $A$ is independent
of the $q^{0}$ direction.
Then following the methods described in sec.II, one can straightforwardly 
obtain the effective description of the above model on an infinite cylinder.
The stability analysis
follows provided the linear terms in the reaction kinetics of the effective 
theory meet the stability condition. The above analysis shows that the thickness
can induce an anisotropy in the speed of the traveling-wave. 
\paragraph*{}
The effective description outline in the paper can be easily extended to any R-D models
like Gierer-Meinhardt model~\cite{gier}, and other R-D models~\cite{murray,biology,philip}. The analysis
may prove useful in the study of the rearrangement of spatial patterns during various
stages of growth in animals. Turing-like models also find applications in diverse
areas like material sciences~\cite{t1}, hydrodynamics~\cite{t2},
 astrophysics~\cite{t3}, etc. In these systems, under certain conditions it is conceivable that the thickness
 and curvature can play a significant role, and hence similar effective descriptions may be suitable.

\section{ACKNOWLEDGEMENT}
\paragraph*{}
I acknowledge Sreedhar Dutta for suggesting the problem, and for extensive discussions. I also
thank him for various helpful suggestions during the preparation of the manuscript.

\bibliography{ref}

\end{document}